\documentclass{article}
\usepackage{amssymb}
\usepackage{amsfonts}
\usepackage{amsmath}

\begin{document}

\begin{center}
\bigskip

\bigskip

\bigskip

\textbf{\Large{Ray  space `Riccati'  evolution and geometric phases for} $N$-%
\textbf{level quantum systems}}

\bigskip

S. Chaturvedi

\textit{School of Physics. University of Hyderabad, Hyderabad 500 046, India.%
}

E. Ercolessi\footnote{\textit{Corresponding author (ercolessi@bo.infn.it).}}

\textit{Physics Dept., University of Bologna, CNISM and INFN, 46 v.Irnerio,
I-40126, Bologna, Italy.}

G. Marmo

\textit{Dipartimento di Scienze Fisiche, University of Napoli and INFN,
v.Cinzia, I-80126, Napoli, Italy.}

G. Morandi

\textit{Physics Dept., University of Bologna, CNISM and INFN, 6/2 v.le Berti
Pichat, I-40127, Bologna, Italy.}

N. Mukunda

\textit{Centre for High Energy Physics, Indian Institute of Science,
Bangalore 560 012, India.}

R. Simon

\textit{The Institute of Mathematical Sciences, C.I.T Campus, Chennai 600
113, India.}

\bigskip

\textbf{Abstract}
\end{center}

We present a simple derivation of the matrix Riccati equations governing the reduced dynamics 
as one descends from the group $\mathbb{U}\left( N\right)$ describing the Schr\"odinger evolution of an $N$-level quantum system to the various coset spaces, Grassmanian manifolds, associated with it. The special case pertaining to the geometric phase in $N$-level systems is described in detail. Further, we  show how the matrix Riccati equation thus obtained can be reformulated
as an equation describing Hamiltonian evolution in a classical phase space and establish correspondences between the two descriptions.


\newpage

\section{Introduction.}

There is currently considerable interest in studying various properties of
quantum systems with finite-dimensional state spaces, for instance in
understanding entanglement in the context of Quantum Information and Quantum
Computation \cite{nielsen}.
Perhaps surprisingly, there seem to be many things yet to be learnt in such
systems. From the standpoint of fundamental Quantum Mechanics as well, there
has been longstanding interest in for example extending the concept of
Wigner distribution to finite dimensions and its role in quantum state estimation \cite{wootters}. In this and other contexts one finds that many new features specific to the number of dimensions appear\cite{Iva}-\cite{BM} that often have no counterpart in the infinite dimensional case .    

For two-level systems, $N=2$, it is well known that the Poincar\'{e}-Bloch
sphere $\mathbb{S}^{2}$ provides an excellent realization and practical tool
for dealing with pure (as well as mixed) quantum states \cite{schleich}. Some efforts to
extend the Poincar\'{e} sphere construction to $N\geq 3$ also exist \cite{khanna}. In
another direction, there has been a systematic programme of `unitary
integration' methods \cite{wei}, which also generalize the Poincar\'{e} sphere concept
to higher dimensions, and lead to methods of projecting quantum dynamics at
the $N$-level Hilbert space governed by a Schr\"{o}dinger equation to
various `base spaces' of lower dimension. From a mathematical point of view, the central 
ideas and structures underlying these `reductions' are similar to those known and extensively developed, over a long period of time, in the context of classical dynamical systems with the 
objective of seeking dynamical superpositions in non linear evolution equations \cite{lie}.  In a recent work of Uskov and Rau \cite{rao}, one
begins with the group $\mathbb{SU}\left( N\right) $ acting on the state
space of an $N$-level system, the Schr\"{o}dinger unitary evolution operator
being also an element of $\mathbb{SU}\left( N\right) $. For any partition: $%
N=n_{1}+n_{2}$, one has the subgroup $\mathbb{SU}\left( n_{1}\right) \times
\mathbb{SU}\left( n_{2}\right) $ and the coset space $\mathbb{SU}\left(
N\right) /\mathbb{SU}\left( n_{1}\right) \times \mathbb{SU}\left(
n_{2}\right) $, which is a Grassmanian manifold and which functions as the
base manifold of a fiber bundle, the total space being $\mathbb{SU}\left(
N\right) $. Using matrices outside of $\mathbb{SU}\left( N\right) $
generated by nilpotent matrices in a well-chosen manner, a parametrization
of the base manifold by a set of $n_{1}\cdot n_{2}$ independent complex
coordinates is set up. It is then shown that the original Schr\"{o}dinger
evolution projects down to a system of matrix Riccati equations for these
base-space coordinates, and some connections to geometric phases 
\cite{geom} are indicated.

The aim of the present work is to give a treatment of this problem of
reduction of Schr\"{o}dinger evolution to various base spaces in a manner
that works throughout within the unitary group $\mathbb{U}\left( N\right) $
intrinsic and natural to Quantum Mechanics, and to obtain the Riccati
equations in a rather elementary manner. Thus the use of `nilpotent
generators' is entirely avoided, and in addition the region of $\mathbb{U}%
\left( N\right) $ covered by the usual complex variables is easily seen. \
Following this, in the case $n_{1}=N-1,n_{2}=1$ appropriate for discussing
evolution of pure states by the Schr\"{o}dinger equation, it is shown that
the base space can be conveniently viewed as a classical phase space, and
the Riccati equations expressing quantum dynamics then appear as classical
Hamiltonian equations with a suitable Hamiltonian function. This framework
is then used to describe in detail the structure of pure state geometric
phases for the original $N$-level system: one sees the extent to which such
phases can be expressed in purely classical terms, and also particularly
clearly that they are ray-space quantities. 

A brief summary of the present work is as follows: In Section 2 we derive the Riccati 
equations associated with various coset spaces of the group $\mathbb{U}(N)$ governing the Schr\"odinger evolution of an $N$ level quantum system without resorting to nilpotent operators. In Section 3 we apply the general formalism of Section 2 to  a special case
appropriate to the geometric phase. Here we also show how the resulting equations can be reformulated in purely classical terms and compare the two perspectives on the geometric phase. Section 4 contains our concluding remarks.        

\bigskip

\section{$N$-Level System Dynamics, Coset Spaces, Riccati Equations.}

Consider an $N$-level quantum system with a Hermitian Hamiltonian matrix $%
H\left( t\right) $ which may be time-dependent. The unitary evolution
operator $U\left( t\right) $ is an element of $\mathbb{G}=\mathbb{U}\left(
N\right) $ obeying the equation:%
\begin{equation}
i\overset{\cdot }{U}\left( t\right) =H\left( t\right) U\left( t\right) ,%
\text{ \ }U\left( t_{0}\right) =\mathbb{I} . \label{muk1}
\end{equation}

When $H$ \ is time-independent, one sees easily that the solutions of this
equation are given by the one-parameter group generated by $iH$ acting on
$\mathbb{G}$ from the left.

The right action of $\mathbb{G}$ on itself acts transitively
on the family of solutions. Therefore, all the solutions are
simply obtained by acting on the subgroup generated by $iH$. This
circumstance  gives rise to the nonlinear superposition rule
present in Lie-Scheffers systems \cite{lie}.

As discussed below, when we consider  `decompositions' of $\mathbb{G}$, we
may relate our equations of motion with other evolution equations
on the space defined by the decomposition, the "reduced space". On
this reduced space the evolution equations are in general
nonlinear and acquire the form of Riccati-type equation.
Many `reduction procedures' are available in the literature, see
for instance \cite{12} for some applications. What we consider here
are specific instances applied to relevant physical situations which arise
in quantum mechanics.

To `decompose' the unitary group $\mathbb{G}$, let us 
consider any partition: $N=n_{1}+n_{2}$ and the subgroup:%
\begin{equation}
\mathbb{H}=\mathbb{U}\left( n_{1}\right) \times \mathbb{U}\left(
n_{2}\right) \subset \mathbb{G},  \label{muk2}
\end{equation}%
with the factor $\mathbb{U}\left( n_{1}\right) $ acting on the first $n_{1}$
dimensions, $\mathbb{U}\left( n_{2}\right) $ on the rest. The idea is to
describe the coset space $\mathbb{G}/\mathbb{H}$ `nicely' and obtain from
Eq. (\ref{muk1}) by `projection' an evolution equation on it. Write a
general matrix in $\mathbb{G}$ in block form as:%
\begin{equation}
\begin{array}{c}
U=\left\vert
\begin{array}{cc}
A & B \\
C & D%
\end{array}%
\right\vert \\
\end{array},
\label{muk3}
\end{equation}
where $A, B, C, D$ are $n_{1}\times n_{1}, n_{1}\times n_{2},n_{2}\times n_{1},n_{2}\times
n_{2}$ dimensional matrices respectively. The condition $U^{\dag }U=\mathbb{I}$ becomes:%
\begin{equation}
A^{\dag }A+C^{\dag }C=\mathbb{I},\text{ \ }A^{\dag }B+C^{\dag }D=0,\text{ \ }%
D^{\dag }D+B^{\dag }B=\mathbb{I}.  \label{muk4}
\end{equation}
Under right multiplication by an element $$\left\vert
\begin{array}{cc}
U_{1} & \mathbf{0} \\
\mathbf{0} & U_{2}%
\end{array}%
\right\vert $$ in $\mathbb{H}$, 
the changes in \ \thinspace $A,B,C,D$ are:
\begin{equation}
A\rightarrow AU_{1},\text{ }B\rightarrow BU_{2},\text{ \ }C\rightarrow
CU_{1},\text{ \ }D\rightarrow DU_{2}.%
\label{muk5}
\end{equation}%
So, in case $A$ and $D$ are non-singular, $CA^{-1}$ and $BD^{-1}$ are
invariant, i.e. they are constants over each coset, and so are $CA^{\dag }$
and $BD^{\dag }$.

Let us now limit ourselves to the subset of $\mathbb{G}$ in which $A,D$ are
both non-singular. Then one can get a nice coset representative as follows:
using the freedom of right multiplication by elements in $\mathbb{H}$, we
can use the polar decomposition of $A$ and $D$ to bring them both to
Hermitian positive-definite forms:
\begin{equation}
\text{\rm{coset representative}}=U_{0}=\left\vert
\begin{array}{cc}
A_{0} & B_{0} \\
C_{0} & D_{0}%
\end{array}%
\right\vert ,\text{ }A_{0}^{\dag }=A_{0}>0,\text{ }D_{0}^{\dag }=D_{0}>0.
\label{muk6}
\end{equation}%
The coset invariant mentioned above can then be defined as an $n_{1}\times n_{2}$ complex rectangular matrix:%
\begin{equation}
Z=B_{0}D_{0}^{-1}.  \label{muk7}
\end{equation}%
We can now use Eq.(\ref{muk4}) to express $A_{0},B_{0},C_{0},D_{0}$ in
terms of $Z$:%
\begin{equation}
\begin{array}{c}
A_{0}=\Gamma _{1}^{-1/2},\text{ }B_{0}=Z\Gamma _{2}^{-1/2},\text{ }%
C_{0}=-Z^{\dag }\Gamma _{1}^{-1/2},\text{ }D_{0}=\Gamma _{2}^{-1/2} 
\end{array},
\label{muk8}
\end{equation}%
where $\Gamma _{1}=\mathbb{I}+ZZ^{\dag }$ and $\text{ }\Gamma _{2}=%
\mathbb{I}+Z^{\dag }Z$ are $n_{1}\times n_{1}$ and $n_{2}\times n_{2}$ positive hermitian matrices intertwined with each other through $Z$:
$\Gamma _{1}Z=Z\Gamma _{2}$.
Thus, in the region of $\mathbb{G}$ defined above, a general matrix is $U$ in 
Eq.(\ref{muk3}) can be written as:%
\begin{equation}
\begin{array}{c}
U=\left\vert
\begin{array}{cc}
A_{0} & B_{0} \\
C_{0} & D_{0}%
\end{array}%
\right\vert \left\vert
\begin{array}{cc}
U_{1} & \mathbf{0} \\
\mathbf{0} & U_{2}%
\end{array}%
\right\vert 
\end{array},
\label{muk9}
\end{equation}%
giving $A=A_{0}U_{1},\text{ }B=B_{0}U_{2},\text{ \ }C=C_{0}U_{1},\text{ \ }%
D=D_{0}U_{2}$.
Using this in the evolution equation (\ref{muk1}) and writing $H\left(
t\right) $ in block form:%
\begin{equation}
H\left( t\right) =\left\vert
\begin{array}{cc}
H_{1}\left( t\right) & V\left( t\right) \\
V\left( t\right) ^{\dag } & H_{2}\left( t\right)%
\end{array}%
\right\vert , \label{muk10}
\end{equation}%
we get:%
\begin{equation}
\begin{array}{c}
i\frac{d}{dt}\left( \left\vert
\begin{array}{cc}
A_{0} & B_{0} \\
C_{0} & D_{0}%
\end{array}%
\right\vert \left\vert
\begin{array}{cc}
U_{1} & \mathbf{0} \\
\mathbf{0} & U_{2}%
\end{array}%
\right\vert \right) =\left\vert
\begin{array}{cc}
H_{1} & V \\
V^{\dag } & H_{2}%
\end{array}%
\right\vert \left\vert
\begin{array}{cc}
A_{0} & B_{0} \\
C_{0} & D_{0}%
\end{array}%
\right\vert \left\vert
\begin{array}{cc}
U_{1} & \mathbf{0} \\
\mathbf{0} & U_{2}%
\end{array}%
\right\vert \\
\end{array},
\nonumber
\end{equation}
i.e.: 
\begin{equation}
i\left\vert
\begin{array}{cc}
\overset{\cdot }{A}_{0} & \overset{\cdot }{B}_{0} \\
\overset{\cdot }{C}_{0} & \overset{\cdot }{D}_{0}%
\end{array}%
\right\vert =\left\vert
\begin{array}{cc}
H_{1} & V \\
V^{\dag } & H_{2}%
\end{array}%
\right\vert \left\vert
\begin{array}{cc}
A_{0} & B_{0} \\
C_{0} & D_{0}%
\end{array}%
\right\vert -i\left\vert
\begin{array}{cc}
A_{0} & B_{0} \\
C_{0} & D_{0}%
\end{array}%
\right\vert \left\vert
\begin{array}{cc}
\overset{\cdot }{U}_{1}U_{1}^{-1} & \mathbf{0} \\
\mathbf{0} & \overset{\cdot }{U}_{2}U_{2}^{-1}%
\end{array}%
\right\vert ,
\nonumber
\end{equation}
which yields
\begin{eqnarray}
i\overset{\cdot }{A}_{0}&=&H_{1}A_{0}+VC_{0}-iA_{0}%
\overset{\cdot }{U}_{1}U_{1}^{-1}, \nonumber\\
\text{\textit{\ }}i\overset{\cdot }{B}_{0}&=&H_{1}B_{0}+VD_{0}-iB_{0}\overset{%
\cdot }{U}_{2}U_{2}^{-1}, \nonumber\\
\text{\textit{\ }}i\overset{\cdot }{C}_{0}&=&V^{\dag }A_{0}+H_{2}C_{0}-iC_{0}%
\overset{\cdot }{U}_{1}U_{1}^{-1}, \\
\text{\textit{\ }}i\overset{\cdot }{D}_{0}&=&V^{\dag }B_{0}+H_{2}D_{0}-iD_{0}%
\overset{\cdot }{U}_{1}U_{1}^{-1}.%
\nonumber
\label{muk11}
\end{eqnarray}
Using the second and fourth of these we get an `autonomous' equation for $Z$:%
\begin{eqnarray}
i\overset{\cdot }{Z}&=&iB_{0}D_{0}^{-1}-iB_{0}D_{0}^{-1}\overset{\cdot }{D}%
D_{0}^{-1}\nonumber \\
&=&H_{1}Z+V-iB_{0}\overset{\cdot }{U}_{2}U_{2}^{-1}D_{0}^{-1}-Z\left( V^{\dag
}Z+H_{2}-iD_{0}\overset{\cdot }{U}_{2}U_{2}^{-1}D_{0}^{-1}\right) \nonumber\\
&=&V+H_{1}Z-ZH_{2}-ZV^{\dag }Z \;,%
\label{muk12}
\end{eqnarray}%
where, for brevity, the $t$-dependencies of $V,H_{1},H_{2},V^{\dag }$ have been omitted.

This is a matrix Riccati equation for evolution on the coset space. The
complete quantum evolution (\ref{muk1}) involves also equations for $\overset{\cdot}{U}_{1}$ and $\overset{\cdot }{U}_{2}$. This derivation seems
simpler and more direct than others \cite{rao}, and shows the appearance of the
Riccati structure rather clearly, including the result of the projection.

\section{Connections to the Geometric Phase.}

To connect up to the Geometric Phase, we limit to the choices $%
n_{1}=N-1,n_{2}=1$. So in the Hamiltonian matrix $H\left( t\right) $ of (\ref%
{muk10}), $H_{1}$ is $\left( N-1\right) \times \left( N-1\right) $, $V$ is
an $\left( N-1\right) $-component column vector, and $H_{2}$ is a single
real quantity. We recall here some notations:

\begin{itemize}

\item $\mathcal{H}^{\left( N\right) }$: complex $N$-dimensional Hilbert space consisting of  vectors $\psi ,\psi ^{\prime }.$
\item $\mathcal{B}_{N}$ : unit sphere in $\mathcal{H}^{\left( N\right)}$ of real dimension $(2N-1)$ identified as a coset space $U\left( N\right) /U\left(
N-1\right) =G/H_{0},\text{ }H_{0}=U\left( N-1\right).$ 
\item $\mathcal{R}_{N}$ :space of unit rays, of real dimension 
$2\left( N-1\right)$ identified as a quotient and coset space $\mathcal{B}_{N}/U\left( 1\right)
\simeq G/H=U\left( N\right) /U\left( N-1\right) \times U\left( 1\right);
H=H_{0}\times U\left( 1\right) =U\left( N-1\right) \times U\left( 1\right)$.%
\end{itemize}

We now consider four aspects:

\bigskip
\noindent
$\left( a\right) $ \ \underline{(Local) coordinates over $\mathcal{R}_{N}$ }%
: 

\medskip
Let us limit ourselves to that part of $\mathcal{H}^{\left( N\right) },%
\mathcal{B}_{N}$ in which the last, $N-th$ component of $\psi \in \mathcal{H}%
^{\left( N\right) }$ is non-zero. Then, for a vector $\psi \in \mathcal{B}%
_{N}$ and its image $\rho \in \mathcal{R}_{N}$ we can say:%
\begin{equation}
\begin{array}{c}
\psi =\frac{1}{\gamma ^{1/2}}e^{i\alpha }\left\vert
\begin{array}{c}
z ,\\
1%
\end{array}%
\right\vert ,\text{ }z\in \mathcal{H}^{\left( N-1\right) },\text{ }\gamma
=1+z^{\dag }z;\text{ }\psi =e^{i\alpha }\psi _{0}\left( z\right), \\
\rho =\psi \psi ^{\dag }=\psi _{0}\left( z\right) \psi _{0}\left( z\right)
^{\dag },\text{ }\psi _{0}\left( z\right) =\frac{1}{\gamma ^{1/2}}\left\vert
\begin{array}{c}
z \\
1%
\end{array}%
\right\vert%
.
\end{array}
\label{muk14}
\end{equation}
So $z\in \mathcal{H}^{\left( N-1\right) }$ becomes a system of coordinates
for the base space $\mathcal{R}_{N}=\mathcal{B}_{N}/U\left( 1\right) \simeq
G/H=U\left( N\right) /U\left( N-1\right) \times U\left( 1\right) $. When
useful we will later write:%
\begin{equation}
z_{r}=q_{r}+ip_{r},\text{ \ }r=1,2,...,N-1 , \label{muk15}
\end{equation}%
so $q_{r},p_{r}$ are $2\left( N-1\right) $ real independent local
coordinates on ray space.

\bigskip

\noindent
$\left( b\right) $ \underline{Schr\"{o}dinger equation on $\mathcal{H}%
^{\left( N\right) }$ to Riccati equation on $\mathcal{R}_{N}$ }

\medskip
Consider the Schr\"{o}dinger equation for a vector $\psi \left( t\right) \in
\mathcal{H}^{\left( N\right) }$:%
\begin{equation}
i\overset{\cdot }{\psi }\left( t\right) =H\left( t\right) \psi \left(
t\right) =\left\vert
\begin{array}{cc}
H_{1} & V \\
V^{\dag } & H_{2}%
\end{array}%
\right\vert \psi \left( t\right).  \label{muk16}
\end{equation}%
Separating the upper $\left( N-1\right) $ components of $\psi $ denoted by $\xi$ from the $N-th$
one $\eta$:%
\begin{equation}
\psi =\left\vert
\begin{array}{c}
\xi \\
\eta%
\end{array}\right\vert ,\text{ \ }
\label{muk17}
\end{equation}
Eq.(\ref{muk16}) may be rewritten as:%
\begin{equation}
i\overset{\cdot }{\xi }=H_{1}\xi +V\eta ,\text{ \ }i\overset{\cdot }{\eta }%
=V^{\dag }\xi +H_{2}\eta . \label{muk18}
\end{equation}%
If we now set: $z=\xi /\eta $, in analogy to Eq.(\ref{muk7}), we get its
evolution equation:%
\begin{eqnarray}
i\overset{\cdot }{z}&=&i\overset{\cdot }{\xi }/\eta -i\xi \overset{\cdot }{%
\eta }/\eta ^{2} \nonumber\\
&=&V+H_{1}z-zH_{2}-zV^{\dag }z .%
\label{muk19}
\end{eqnarray}%
This is again a Riccati equation, like (\ref{muk12}), but now $z$ is an $%
\left( N-1\right) $-component column vector, as is $V$, and $H_{2}$ is a
single real variable. Now we have obtained the Riccati structure for a
single solution of the Schr\"{o}dinger equation, not
using the entire unitary evolution operator. So the quantum-mechanical ray
space evolution is given by the non-linear Riccati equation (\ref{muk19}).

\bigskip
\noindent
$\left( c\right) $ \underline{Reformulation in classical phase space form}.

\medskip
We now show that Eq.(\ref{muk19}) can be written in completely classical
form on $\mathcal{R}_{N}$ \underline{regarded as a phase space}. We start
from Eq.(\ref{muk14}) and define first a one-form $\theta _{0}$ and then a
symplectic two-form $\omega _{0}$ on $\mathcal{R}_{N}$ as follows:%
\begin{eqnarray}
\theta _{0}&=&-i\psi _{0}\left( z\right) ^{\dag }d\psi _{0}\left( z\right)\nonumber\\ &=&-%
\frac{i}{\sqrt{\gamma }}\left\vert
\begin{array}{cc}
z^{\dag } & 1%
\end{array}%
\right\vert \left\{ \gamma ^{-1/2}\left\vert
\begin{array}{c}
dz \\
0%
\end{array}%
\right\vert +\left\vert
\begin{array}{c}
z \\
1%
\end{array}%
\right\vert d\gamma ^{-1/2}\right\} \nonumber\\
&=&\frac{1}{\gamma }\rm{Im}~z^{\dag }dz \nonumber\\
&=&
\frac{1}{\gamma }\left( q_{r}dp_{r}-p_{r}dq_{r}\right),
\label{muk20}
\end{eqnarray}
and:%
\begin{eqnarray}
\omega _{0}&=&d\theta _{0}=-id\psi _{0}\left( z\right) ^{\dag }\wedge d\psi
_{0}\left( z\right)\nonumber \\
&=&\frac{2}{\gamma }dq_{r}\wedge dp_{r}-\frac{2}{\gamma ^{2}}\left(
q_{r}dq_{r}+p_{r}dp_{r}\right) \wedge \left( q_{s}dp_{s}-p_{s}dq_{s}\right)
\nonumber\\
&=&2\left\vert
\begin{array}{cc}
dq_{r} & dp_{r}%
\end{array}%
\right\vert \wedge\left\vert
\begin{array}{cc}
L_{rs} & M_{rs} \\
-M_{rs} & L_{rs}%
\end{array}%
\right\vert \left\vert
\begin{array}{c}
dq_{s} \\
dp_{s}%
\end{array}%
\right\vert, 
\label{muk21}
\end{eqnarray}
where
\begin{eqnarray}
L_{rs}&=&-L_{sr}=\left( q_{r}p_{s}-q_{s}p_{r}\right) /2\gamma ^{2} ,\nonumber\\
M_{rs}&=&M_{sr}=\frac{1}{2\gamma }\delta _{rs}-\left(
q_{r}q_{s}+p_{r}p_{s}\right) /2\gamma ^{2}.%
\end{eqnarray}
The two-form $\omega _{0}$ is indeed non-singular, and the inverse of the $2\left(
N-1\right) \times 2\left( N-1\right) $ matrix above can be computed:%
\begin{equation}
\begin{array}{c}
\left\vert
\begin{array}{cc}
\mathbf{L} & \mathbf{M} \\
-\mathbf{M} & \mathbf{L}%
\end{array}%
\right\vert ^{-1}=\left\vert
\begin{array}{cc}
\mathbf{X} & \mathbf{Y} \\
-\mathbf{Y} & \mathbf{X}%
\end{array}%
\right\vert ,\\
\end{array}
\label{muk22}
\end{equation}%
where 
\begin{eqnarray}
X_{rs}&=&-X_{sr}=2\gamma \left( q_{r}p_{s}-q_{s}p_{r}\right), \nonumber\\
Y_{rs}&=&Y_{sr}=-2\gamma \left( \delta _{rs}+q_{r}q_{s}+p_{r}p_{s}\right),%
\end{eqnarray}
and we can read off the fundamental $PB$'s on ray space in $q-p$ and $%
z-z^{\ast }$ form:%
\begin{eqnarray}
\left\{ q_{r},q_{s}\right\} &=&\left\{ p_{r},p_{s}\right\} =\gamma \left(
q_{r}p_{s}-q_{s}p_{r}\right)/2 =X_{rs}/4 ,\nonumber\\
\left\{ q_{r},p_{s}\right\} &=&-\gamma \left( \delta
_{rs}+q_{r}q_{s}+p_{r}p_{s}\right)/2 =Y_{rs}/4, \nonumber\\
\left\{ z_{r},z_{s}^{\ast }\right\} &=&i\gamma \left( \delta
_{rs}+z_{r}z_{s}^{\ast }\right) ,\text{ \ }\left\{ z_{r},z_{s}\right\}
=\left\{ z_{r}^{\ast },z_{s}^{\ast }\right\} =0.%
\label{muk23}
\end{eqnarray}%
From here the following useful $PB$'s can be obtained:%
\begin{eqnarray}
\left\{ z_{r},\gamma \right\} &=&i\gamma ^{2}z_{r},\text{ }\left\{
z_{r},\gamma ^{-1}\right\} =-iz_{r} ,\nonumber\\
\left\{ z_{r},z_{s}^{\ast }/\gamma \right\} &=&i\delta _{rs},\text{ }\left\{
z_{r},f\left( z\right) /\gamma \right\} =-iz_{r}f\left( z\right).%
\label{muk24}
\end{eqnarray}%
Here $f\left( z\right) $ is \ analytic in the $z_{r}$'s. Using all this, we
can show that the quantum-mechanical Riccati evolution equation (\ref{muk19}%
) can be written as a purely classical Hamiltonian evolution on $\mathcal{R}%
_{N}$ with the $PB$'s given in (\ref{muk23}):%
\begin{eqnarray}
\overset{\cdot }{z}&=&-i\left( V+H_{1}z-zH_{2}-zV^{\dag }z\right)\nonumber\\ 
&=&\left\{ z,%
\mathcal{H}\left( z,z^{\dag }\right) \right\},
\end{eqnarray} 
where
\begin{eqnarray}
\mathcal{H}\left( z,z^{\dag }\right) &=&-\left( H_{2}+z^{\dag }V+V^{\dag
}z+z^{\dag }H_{1}z\right) /\gamma \nonumber\\
&=&-\frac{1}{\gamma }\left\vert
\begin{array}{cc}
z^{\dag } & 1%
\end{array}%
\right\vert \left\vert
\begin{array}{cc}
H_{1} & V \\
V^{\dag } & H_{2}%
\end{array}%
\right\vert \left\vert
\begin{array}{c}
z \\
1%
\end{array}%
\right\vert 
\nonumber\\
&=&-\psi _{0}\left( z\right) ^{\dag }H\psi _{0}\left(z\right).
\label{muk25}
\end{eqnarray}

The motion of $z$ in $\mathcal{R}_{N}$, induced by unitary Schr\"{o}%
dinger evolution in $\mathcal{H}^{\left( N\right) }$, is thus a continuous
classical canonical transformation.

\bigskip

\noindent
$\left( d\right) $ \underline{Geometric phase and its `classical' aspects }

\medskip

We consider Schr%
\"{o}dinger evolution as given by Eq.(\ref{muk16}), with no appeal to the adiabatic approximation. The Riccati equation (\ref{muk19}) for $z$, as we have seen, is a 
consequence of Eq.(\ref{muk16}), and from it we find for $\gamma$:%
\begin{equation}
\begin{array}{c}
\overset{\cdot }{\gamma }\left( t\right) =i\gamma \left( t\right) \left(
V^{\dag }z\left( t\right) -z\left( t\right) ^{\dag }V\right) ,\text{ \ }%

\end{array}
\label{muk26}
\end{equation}%
where in general $V=V\left( t\right)$. Comparing Eqns.(\ref{muk14}) and (\ref{muk17}) gives:
\begin{equation}
\xi =e^{i\alpha }z/\sqrt{\gamma },\text{ \ }\eta =e^{i\alpha }/\sqrt{\gamma }.
\label{muk27}
\end{equation}%
Using these and (\ref{muk26}) in the $\overset{\cdot }{\eta }$ equation of motion in
(\ref{muk18}) gives the evolution equation for the phase $\alpha \left(
t\right) $:%
\begin{equation}
\begin{array}{c}
\overset{\cdot }{\alpha }=-H_{2}-\frac{1}{2}\left( V^{\dag }z+z^{\dag}V\right)%
\end{array}.
\label{muk28}
\end{equation}%
( Both Eqs.(\ref{muk26})(\ref{muk28}) appear in \cite{rao}). In summary, the Schr\"{o}dinger equation (\ref{muk16}) for $\psi \left(
t\right)$ amounts to an autonomous Riccati equation (\ref%
{muk19})for $z$, re-expressed in (\ref{muk24}) in
classical phase space form and equation of motion (\ref{muk28}) for $\alpha \left( t\right)$ where the right-hand side is $\alpha$ independent.

If now $\psi \left( t\right) $ is any (not necessarily cyclic) solution of
the Schr\"{o}dinger equation (\ref{muk16}) between the given 
times $t_{1}$ and $t_{2}$, from general theory  it is known that the geometric phase is the 
difference of two terms, a total and a dynamical phase:
\begin{eqnarray}
\varphi _{geom}&=&\varphi _{tot}-\varphi _{dyn},\nonumber\\
\varphi _{tot}&=&\arg \left( \psi \left( t_{1}\right) ,\psi \left(
t_{2}\right) \right)\nonumber\\
&=&\alpha \left( t_{2}\right) -\alpha \left( t_{1}\right)
+\arg \left( 1+z^{\dag }\left( t_{1}\right) z\left( t_{2}\right) \right),\nonumber \\
\varphi _{dyn}&=&\rm{Im}\int\limits_{t_{1}}^{t_{2}}dt\psi \left( t\right)
^{\dag }\frac{d\psi \left( t\right) }{dt} \nonumber\\
&=&-\int\limits_{t_{1}}^{t_{2}}dt\psi _{0}\left( z\left( t\right) \right)
^{\dag }H\left( t\right) \psi _{0}\left( t\right) \nonumber\\
&=&\int\limits_{t_{1}}^{t_{2}}dt%
\mathcal{H}\left( z\left( t\right) ,z\left( t\right) ^{\ast }\right). 
 \label{muk31}
\end{eqnarray}
The first part is calculable using (\ref{muk28}) and depends only on ray space quantities; the second part is the time integral of the classical Hamiltonian along the ray space trajectory $z(t)$. Thus the geometric phase is seen to involve only ray space quantities, as it must, and is expressed as far as possible in terms of the classical Hamiltonian $
\mathcal{H}\left( z,z^{\ast }\right) $ via the dynamical phase. In special case of 
cyclic evolution, $z\left( t_{1}\right) =z\left( t_{2}\right) $, the
total phase simplifies to just $\alpha \left( t_{2}\right) -\alpha \left(
t_{1}\right) $.

If we adopt the kinematic approach in which there is no use of the Schr\"{o}dinger equation, the results our similar except for a shifting of terms. We now consider directly a (smooth) parametrized curve $\mathcal{C}$ in $%
\mathcal{B}_{N}$ with image $C$ in $\mathcal{R}_{N}$:%
\begin{equation}
\begin{array}{c}
\mathcal{C}=\left\{ \psi \left( s\right) =e^{i\alpha \left( s\right) }\psi
_{0}\left( z\left( s\right) \right) |s_{1}\leq s\leq s_{2}\right\} \subset
\mathcal{B}_{N} ,\\
C=\left\{ \rho \left( s\right) =\psi _{0}\left( z\left( s\right) \right)
\psi _{0}\left( z\left( s\right) \right) ^{\dag }|s_{1}\leq s\leq
s_{2}\right\} \subset \mathcal{R}_{N}\;.%
\end{array}
\label{muk32}
\end{equation}%
The latter is essentially a curve in the space of $z$:%
\begin{equation}
C=\left\{ z\left( s\right) |s_{1}\leq s\leq s_{2}\right\} . \label{muk33}
\end{equation}
Then the geometric phase is again the difference is as usual:
\begin{eqnarray}
\varphi _{geom}\left[ C\right] &=&\varphi _{tot}\left[ \mathcal{C}\right]
-\varphi _{dyn}\left[ \mathcal{C}\right] ,\nonumber \\
\varphi _{tot}\left[ \mathcal{C}\right] &=&\arg \left( \psi \left(
s_{1}\right) ,\psi \left( s_{2}\right) \right) \nonumber\\
&=&\alpha \left( s_{2}\right)
-\alpha \left( s_{1}\right) +\arg \left( 1+z\left( s_{1}\right) ^{\dag
}z\left( s_{2}\right) \right), \nonumber \\
\varphi _{dyn}\left[ \mathcal{C}\right] &=&\rm{Im}\int
\limits_{s_{1}}^{s_{2}}ds\psi \left( s\right) ^{\dag }\frac{d}{ds}\psi
\left( s\right)\nonumber
\\ &=&\rm{Im}\int\limits_{s_{1}}^{s_{2}}ds\psi \left( s\right)
^{\dag }\left\{ i\overset{\cdot }{\alpha }\left( s\right) \psi \left(
s\right) +e^{i\alpha \left( s\right) }\frac{d}{ds}\psi _{0}\left( z\left(
s\right) \right) \right\} \nonumber \\
&=&\alpha \left( s_{2}\right) -\alpha \left( s_{1}\right) +\rm{Im}
\int\limits_{ \stackrel{\scriptstyle{ s_{1}}}{ {\rm along~}C }  }  ^{s_{2}} \psi _{0}\left(
z\left( s\right) \right) ^{\dag }d\psi _{0}\left( z\left( s\right) \right)\nonumber \\ 
&=&
\alpha \left( s_{2}\right) -\alpha \left( s_{1}\right)
+\int\limits_{C}\theta _{0}\;. 
\label{muk34}
\end{eqnarray}
Here we used Eq.(\ref{muk20}). Therefore again we have a ray space quantity:
\begin{equation}
 \varphi _{geom}\left[ C\right] =\arg \left( 1+z\left(
s_{1}\right) ^{\dag }z\left( s_{2}\right) \right) -\int\limits_{C}\theta
_{0}.
\end{equation}
If $C$ is a
closed loop, then $z\left( s_{1}\right) =z\left( s_{2}\right) $, the geometric phase is a purely classical
"symplectic area":%
\begin{equation}
\partial C=0:
\varphi _{geom}\left[ C\right] =-\int\limits_{C}\theta
_{0}=-\iint\limits_{S}\omega _{0},
\label{muk35}
\end{equation}
where $S$ is any two-surface with $\partial S=C$.

\section{Concluding Remarks.}

In the present work, we have  attempted has to highlight the connections between the
following in the quantum evolution of $N$-level systems:

\begin{enumerate}
\item Appearance of Riccati equations in as direct a way as possible to
describe `unitary' evolution in the coset space $\mathbb{G}/\mathbb{U}\left(
n_{1}\right) \times \mathbb{U}\left( n_{2}\right) $ where $\mathbb{G=U}%
\left( N\right) $ and $n_{1}+n_{2}=N$;

\item In the case $n_{1}=N-1,n_{2}=1$: to recast the Schr\"{o}dinger
equation for $\psi $ in $\mathcal{H}^{\left( N\right) \text{ }}$as a Riccati
equation on the ray space $\mathcal{R}_{N}$ plus an equation (\ref{muk28})
for the overall phase $\alpha $ `driven' by this ray-space evolution; to
express the ray-space evolution as a classical Hamiltonian evolution with $PB
$'s (\ref{muk23}) and Hamiltonian function $\mathcal{H}\left( z\right) $; to
study the structure of the $GP$ in both Schr\"{o}dinger evolution and
kinematic case, bring in the classical symplectic structure as far as
possible, and show that the $GP$ is always a ray-space quantity.
\end{enumerate}

We hope the approach to Riccati equations developed in the present work, owing to simplicity and directness, will find useful applications in problems involving `reductions' as illustrated here with geometric phases as an example. It will be interesting to consider \ the situation in which $\mathbb{G}$ is the unitary group associated with a composite Hilbert space of
dimension $N=nm$, with the Hilbert spaces of the subsystems being
of dimension $n$ and $m$ respectively. The subgroup $\mathbb{H}$ will be the tensor product of
$\mathbb{U}\left( n\right)  $ and $\mathbb{U}\left(  m\right)  $,
the so-called group of local transformations. In this way the
reduced space will be related to the set of entangled states \ of
the composite system and this may shed new light on the
classification problem of separable versus entangled states.


\begin{thebibliography}{99}

\bibitem{nielsen} M. Nielsen and I. Chuang, {\it Quantum Computation and Quantum Information}
(Cambridge University Press, New York, 2000); I. Bengtsson and K. $\dot{{\rm Z}}$yczkowski, 
{\it Geometry of Quantum States : An Introduction to Quantum Entanglement}
(Cambridge University Press, New York, 2006). 

\bibitem{wootters} K. S. Gibbons, M. J. Hoffman  and  W. K. Wootters, 
Phys. Rev. A{\bf 70}, 062101 (2004) ; G. S. Agarwal, Phys. Rev. A {\bf 24}, 2889 (1981); M.Ruzzi and D. Galetti, J. Phys. A {\bf 33}, 1065 (1999), N. Mukunda, S. Chaturvedi and R. Simon, Phys. Lett. A {\bf 321} 160 (2004); S. Chaturvedi, E. Ercolessi, G. Marmo, G. Morandi, N. Mukunda and R. Simon, Pramana,J. Phys. {\bf 65}, 981 (2005); A. Vourdas, Rep. Prog. Phys. {\bf 67}, 267(2004); D. Gross, J. Math. Phys. {\bf 47}, 122107 (2006).   

\bibitem{Iva} I. D. Ivanovic, {\it  J. Phys. A} {\bf 14} 3241(1981); W. K. Wootters 
 and  B. D. Fields,  {\it Ann. Phys. (N.Y.)} {\bf 191} 363 (1989); A. R. Calderbank 
, P. J. Cameron, W. M. Kantor and J. J. Seidel, Proc. London. Math. 
Soc. {\bf  75} 436(1997); S. Bandyopadhyay, P. O. Boykin, V. Roychowdhury and F. Vatan, 
 Algorirhmica, {\bf 34} 512 (2002) ; J. Lawrence, C. Brukner and A. Zeilinger, 
 Phys. Rev. A {\bf 65} 032320 (2002); S. Chaturvedi,  
Phys. Rev. A {\bf 65} 044301 (2002) ; A. O. Pittenger and M. H. Rubin, 
 Linear Alg. Appl. {\bf 390} 255 (2004) ; A. O. Pittenger and  M. H. Rubin, 
 J. Phys. A {\bf 38} 6005 (2005); A. Klappenecker and M. R\"otteler, Lecture Notes in Computer Science  {\bf 2948} 137 (2004); K. R. Parthasarathy, Anal. Quantum Probab. Relat. Top. {\bf 7}, 607 (2004).

\bibitem{SP} M. Saniga, M. Planat and H. Rosu, J. Opt.  Quantum 
Semiclass. B{\bf 6}, L19 (2004) ; P. Wocjan and T. Beth, Quantum Information 
and Computation {\bf 5}, 93 (2005); A. Hayashi A, M. Horibe M and Hashimoto T Phys. Rev. A 71, 052331 (2005); I. Bengtsson, AIP Conf. Proceedings {\bf 750}, 63 (2005) ; H. Barnum, Preprint, quant-ph/0205155 (2002);  A. Klappenecker and M. R\"otteler Proc. 2005 IEEE International Symposium on Information Theory, Adelaide, Australia, pp. 1740-1744, 2005; G. Zauner  {\it Quantumdesigns: Grundz\"uge einer nichtkommutativen Designtheorie} Ph.D. thesis (Universit\"at Wien) (1999) 

\bibitem{BM} P. Bianucci, C. Miquel, J. P. Paz and M. Saraceno, Phys. Lett. A {\bf 297}, 353 (2002); C. Miquel, J. P. Paz, M. Saraceno, E. Knill, R. Laflamme and C. Negrevergne, Nature (London) {\bf 418}, 59 (2002); J. P. Paz, A. J. Roncaglia  and M. Saraceno M, Phys. Rev. A 72, 012309 (2005);  M. Horibe, A. Takami, T. Hashimoto and A. Hayashi,  Phys. Rev. A {\bf 65}, 032105 (2002).

\bibitem{schleich} See for instance, W.P. Schleich, 
{\it Quantum Optics in Phase Space} (Wiley-VCH, Weinheim, 2001)

\bibitem{khanna} G. Khanna, S. Mukhopadhyay, R. Simon, and N. Mukunda, J. Phys. A {253}, 55 (1997); Arvind, K . S. Mallesh and N. Mukunda, J. Phys. A {30}, 2417 (1997); L. Jak$\acute{{\rm o}}$biczyk and M. Sienicki, Phys. Lett. A{\bf 286}, 383 (2001); G. Kimura Phys. Lett. A{\bf 314}, 3339 (2003); M. S. Byrd and N. Khaneja, Phys. Rev. A{\bf 68}, 062322 (2003).
\bibitem{wei} See, for instance, J. Wei and E. Norman, J. Math.Phys.,{\bf 4}, 575 (2001); 
G. Dattoli, J. C. Gallardo, and A. Torre, Riv. Nuovo. Cimento {\bf 11}, 1 (1988); B. A. Shadwick and W. F. Buell, Phys. Rev. Lett. {\bf 79}, 5189 (1997). 

\bibitem{lie} See, for instance, J. F. Cari$\tilde{{\rm n}}$ena, J. Grabowski and G. Marmo, 
{\it Lie-Scheffers Systems: A Geometric Approach}, ( Bibliopolis, Naples, 2000); J. Grabowski, G. Landi, G. Marmo and G. Vilasi, Fortsch. Phys. {\bf 46}, 393 (1994).
\bibitem{rao} D. B. Uskov and A. R. P. Rao, Phys. Rev. A{\bf 74}, 030304(R) (2006); See also, A. R. P. Rau, Phys. Rev. Lett. {\bf 81}, 4785 (1998); A. R. P. Rao and Weichang Zhao, Phys. Rev. A{\bf 71}, 063822 (2005); A. R. P. Rau, G. Selvaraj and D. B. Uskov, Phys. Rev. A{\bf 71}, 062316 (2005). 
\bibitem{geom} S. Pancharatnam, Proc. Indian Acad. Sci. Section {\bf 44}, 247 (1956);M. V. Berry, Proc. Roy. Soc. A{\bf 392}, 45 (1984); Y. Aharanov and J. Anandan, Phys. Rev. Lett. {\bf 58}, 1593 (1987); J. Samuel and R. Bhandari, Phys. Rev. Lett. {\bf 60}, 
2339 (1988); N. Mukunda and R. Simon Ann. Phys. (NY) {\bf 228}, 205 
(1993); {\it ibid} {\bf 228}, 269 (1993). Many of the early papers on geometric phase have been 
reprinted in  {\it Geometric Phases in 
Physics} by A. Shapere and F. Wilczek,, (World Scientific, Singapore, 
1989) and in {\it Fundamentals of Quantum Optics}, SPIE Milestone 
Series, edited by G. S. Agarwal, (SPIE Press, Bellington, 1995). For off diagonal geometric phase and that associated with mixed states see, for instance, N. Mukunda, Arvind, S. Chaturvedi and R. Simon,Phys.
Rev. A {\bf 65}, 012102 (2003); S. Chaturvedi, E. Ercolessi, G. Marmo, G. Morandi, N. Mukunda and R. Simon, Eur. Phys. J. C{\bf 35}, 413 (2004); S. Filipp and E. Sj\"oqvist  Phys Rev Lett. {\bf 90}, 050403 (2003) and references cited therein. 

\bibitem{12} G.Marmo, E.J.Saletan, A.Simoni and B.Vitale: \textit{Dynamical
Systems. A Differential Geometric Approach to Symmetry and
Reduction. } J.Wiley, 1985; A.M.Perelomov: \textit{Integrable
Systems of Classical Mechanics and Lie Algebras. }Birkh\"{a}user,
1990.
 
 


\end{thebibliography}
\end{document}